\documentclass[aps,prb,twocolumn,showpacs,superscriptaddress]{revtex4}
\usepackage{amsmath}
\usepackage{amssymb}
\usepackage{epsfig}
\usepackage{color}
\usepackage{graphicx,amsmath}
\usepackage{hyperref}
\hypersetup{colorlinks,citecolor=blue,linkcolor=blue,urlcolor=blue}

\begin{document}

\title{The influence of topological phase transition
on the superfluid density of overdoped copper oxides}
\author{V. R. Shaginyan}\email{vrshag@thd.pnpi.spb.ru}
\affiliation{Petersburg Nuclear Physics Institute, NRC
"Kurchatov Institute", Gatchina, 188300,
Russia}\affiliation{Clark Atlanta University, Atlanta, GA 30314,
USA}\author{V. A. Stephanovich}\affiliation{Institute of
Physics, Opole University, Opole, 45-052, Poland} \author{A. Z.
Msezane}\affiliation{Clark Atlanta University, Atlanta, GA
30314, USA}\author{G. S. Japaridze}\affiliation{Clark Atlanta
University, Atlanta, GA 30314, USA}\author{K. G.
Popov}\affiliation{Komi Science Center, Ural Division, RAS,
Syktyvkar, 167982, Russia}

\begin{abstract}

We show that a topological quantum phase transition, generating
flat bands and altering Fermi surface topology, is a primary
reason for the exotic behavior of the overdoped high-temperature
superconductors represented by $\rm La_{2-x}Sr_xCuO_4$, whose
superconductivity features differ from what is described by the
classical Bardeen-Cooper-Schrieffer theory [J.I. Bo\^zovi\'c, X.
He, J. Wu, and A. T. Bollinger, Nature {\bf 536}, 309 (2016)].
We demonstrate that 1) at temperature $T=0$, the superfluid
density $n_s$ turns out to be considerably smaller than the
total electron density; 2) the critical temperature $T_c$ is
controlled by $n_s$ rather than by doping, and is a linear
function of the $n_s$; 3) at $T>T_c$ the resistivity $\rho(T)$
varies linearly with temperature, $\rho(T)\propto \alpha T$,
where $\alpha$ diminishes with $T_c\to 0$, while in the normal
overdoped (non superconducting) region with $T_c=0$, the
resistivity becomes $\rho(T)\propto T^2$. The theoretical
results presented are in good agreement with recent experimental
observations, closing the colossal gap between these empirical
findings and Bardeen-Cooper-Schrieffer-like theories.

\end{abstract}


\maketitle

\section{Introduction}

By now, overdoped copper oxides are realized as simple HTSC,
whose strongly correlated physics can be captured by  the
conventional Bardeen–Cooper– Schrieffer theory (BCS), while
recent experimental studies of overdoped high-$T_c$
superconductors (HTSC) $\rm La_{2-x}Sr_xCuO_4$ discovered strong
deviations of their physical properties from those predicted by
BCS theory \cite{bosovic,zaanen}. These deviations were
surprisingly similar for numerous HTSC samples
\cite{bosovic,zaanen,uemura,uem_n,bern,rour}. The measurements
of the absolute values of the magnetic penetration depth
$\lambda$ and the phase stiffness $\rho_s=A/\lambda^2$ were
carried out on thousands of perfect two dimensional (2D) samples
of $\rm La_{2-x}Sr_xCuO_4$ (LSCO) as a function of the doping
$x$ and temperature $T$. Here $A=d/4k_Be^2$ where $d$ is the
film thickness, $k_B$ is Boltzmann constant, and $e$ is the
electron charge. It has been observed that the dependence of
zero-temperature superfluid density (the density of
superconductive electrons) $n_s=4\rho_s k_Bm^*$ ($m^*$ is the
electron effective mass), is proportional to the critical
temperature $T_c$ over a wide doping range. This dependence
coincides with pervious measurements, and is incompatible with
the standard BCS description. Moreover, $n_s$ turns out to be
considerably smaller than the BCS density $n_{el}$ of
superconductive electrons
\cite{bosovic,zaanen,uemura,uem_n,bern,rour}, which is
approximately equal to the total electron density
\cite{bardeen}. These observations representing the intrinsic
LSCO properties provide unique opportunities for checking and
expanding our understanding of the physical mechanisms
responsible for high-$T_c$ superconductivity. We note, that that
knowing the responsible mechanism can open avenue for chemical
preparation of high-$T_c$ materials with $T_c$ as high as room
temperature
\cite{khs,amusia:2015,volovik:91,volovik,volovik:2015}.

Here we show that the physical mechanism, responsible for above
non-BCS behavior of overdoped LSCO, stems from the topological
fermion condensation quantum phase transition (FCQPT)
accompanied by so-called fermion condensation (FC) phenomenon
generating flat bands
\cite{amusia:2015,khs,volovik:91,volovik,volovik:2015,khodel:1994,shagrep}.
We note that flat bands and extended saddle point singularity
play important role in the theory of HTSC, see e.g. Refs.
\cite{volovik:2015,khodel:1994,shagrep,abrikosov,abrikosov1}.

In order to make our analysis of overdoped LSCO obvious, we use
the model of homogeneous heavy-electron liquid
\cite{shagrep,amusia:2015}. The main experimental facts of Refs.
\cite{bosovic,zaanen} represent vivid qualitative deviations
from those predicted by the classical BCS theory, therefore as a
first step, we can confine ourself to obtaining transparent
analytical results describing quantitatively experimental facts.
Our analysis shows that despite drastic microscopic diversity of
strongly correlated Fermi systems, they exhibit similar behavior
close to FC quantum phase transition point. This is actually
related to the altering of Fermi surface topology during FCQPT.
We emphasize that the quantum physics of all seemingly different
strongly correlated Fermi systems (and overdoped HTSC among
them) is universal and emerges regardless of their underlying
microscopic details like the symmetries of their crystal
lattices. Because we deal effectively with momenta transfers
that are small compared to those of the order of the reciprocal
lattice length (Brillouin zone boundaries), whose contributions
have no effect on the topological properties of the systems
under consideration
\cite{shagrep,amusia:2015,volovik:91,volovik,volovik:2015}. Note
that despite the highly anisotropic electronic band dispersion
in overdoped cuprate HTSC and hence their Fermi surface, our
theory still applies for this case. The point here is that after
FCQPT the Fermi surface, regardless its initial anisotropy,
changes its {\em{topological class}}, thus generating all
aforementioned salient experimentally observed features,
inherent in the fermion condensation state. In other words, any
initially (highly) anisotropic Fermi surface is still
{\em{homotopic}} to simply spherical one as they can be reduced
to each other by continuous deformation \cite{dnf,vm77}. In the
superconducting state, to the first approximation different
regions with the maximal absolute value of the $d$-wave
superconducting order parameter are disconnected. Therefore, the
order parameter can be either even, or odd with respect to a
$\pi/2$ rotation in the ab-plane \cite{abrikosov,abrikosov1}.
Thus, as a first step, we also neglect the d-wave symmetry of
the superconducting order parameter and use the s-wave one.

In our paper, using formalism accounting for the FCQPT, we
investigate overdoped LSCO and show that as soon as the doping
$x$ reaches its FCQPT critical value $x_c$, the features of the
emergent superconductivity begin to differ from those of BCS
theory, as it is predicted long before the experimental
observations are obtained \cite{bosovic,zaanen,qp2}. We
demonstrate that: i) at $T=0$, the superfluid density $n_s$
turns out to be a small fraction of the total density of
electrons; ii) the critical temperature $T_c$ is controlled by
$n_s$ rather than by doping, and is a linear function of the
$n_{s}$. Since FCQPT generates flat electronic bands
\cite{amusia:2015,khs,volovik:91,volovik,volovik:2015}, the
system under consideration exhibits non-Fermi liquid (NFL)
behavior and the resistivity $\rho(T)$ varies linearly with
temperature, $\rho(T)\propto \alpha T$. Since at $x\to x_c$
$\alpha$ diminishes with $T_c$ decreasing,  the system exhibits
Landau Fermi liquid (LFL) behavior at $x>x_c$ and at low
temperatures. These results are in good agreement with recent
experimental observations \cite{bosovic,zaanen,pagl}.

\section{Two-component system}

An important problem for the condensed matter theory is the
explanation of the NFL behavior observed in HTSC beyond critical
point where the low-temperature density of states $N(T\to 0)$
diverges which can generate flat bands without breaking any
ground state symmetry, see e.g. Refs.
\cite{bosovic,volovik:2015,khodel:1994,shagrep,pagl,khod:2015,lifshitz}.
In a homogeneous matter, such a divergence is associated with
the onset of a topological transition at $x=x_c$ signaled by the
emergence of an inflection point at $p=p_F$
\cite{khodel:1994,shagrep,ybalb}
\begin{eqnarray}\label{top}
\varepsilon-\mu&\simeq&-(p_F-p)^2,\ p<p_F,\\
\varepsilon-\mu&\simeq&\,\,\,(p-p_F)^2,\ p>p_F,\nonumber
\end{eqnarray}
at which the electron effective mass diverges as $m^*(T\to
0)\propto T^{-1/2}$,  where $\varepsilon$ is the single -
electron energy spectrum, $p$ is a momentum, $p_F$ is Fermi
momentum and $\mu$ is the chemical potential. Accordingly, at $
x\to x_c$ the density of states diverges
\begin{equation}
N(T\to 0)\propto |\varepsilon-\mu|^{-1/2}. \label{dt}
\end{equation}
As a result, both FC state and the corresponding flat bands
emerge beyond the topological FCQPT
\cite{khod:2015,khodel:1994,shagrep,amusia:2015}, while the
critical temperature turns out to be $T_c\propto \sqrt{x-x_c}$
\cite{abrikosov,abrikosov1}. These results are consistent with
the experimental data  \cite{bosovic}. The detailed
consideration of this case will be published elsewhere.

At $T=0$, the onset of FC in homogeneous matter is attributed to
a nontrivial solution $n_0(p)$ of the variational equation
\cite{khs}
\begin{equation}
\frac{\delta E[n(p)]}{\delta n(p)}-\mu=0 , \ p\in [p_i,p_f],
\label{var}
\end{equation}
where $E$ is a ground state energy functional (its variation
gives a single - electron spectrum $\varepsilon$) and $p_i$,
$p_f$ stand for initial and final momenta, where the solution of
Eq. \eqref{var} exists, see Refs. \cite{amusia:2015,shagrep,
khs} for details.To be more specific, Eq.~\eqref{var} describes
a flat band pinned to the Fermi surface and related to FC.

\begin{figure} [! ht]
\begin{center}
\includegraphics [width=0.44\textwidth]{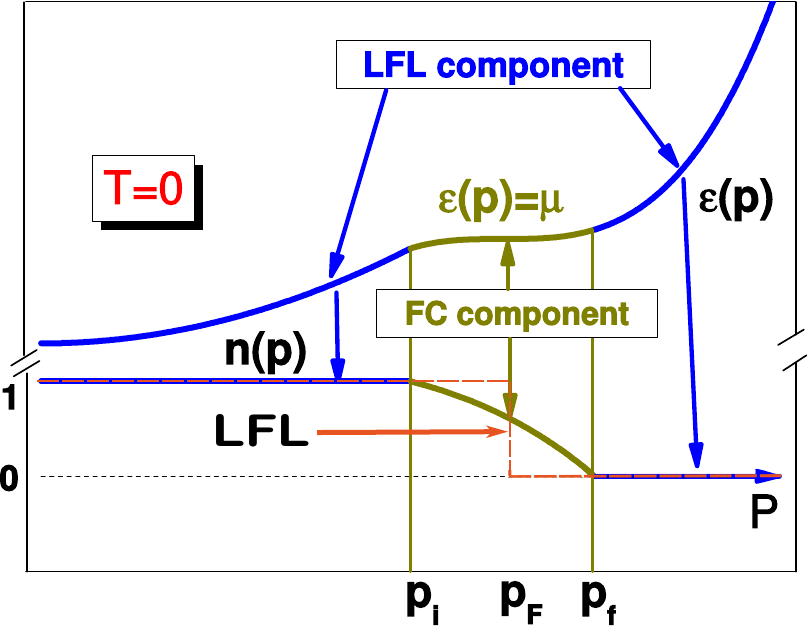}
\end{center}
\caption {(Color online) Schematic plot of two-component
electron liquid at $T=0$ with FC. Red dashed line (marked "LFL")
shows $n(p)$ for the system without FC, which has ordinary step
function shape. Due to the presence of FC, the system is
separated into two components. The first one is a normal Fermi
liquid with the quasiparticle distribution function
$n_0(p<p_i)=1$, and $n_0(p>p_f)=0$. The second one is FC with
$0<n_0(p_i<p<p_f)<1$ and the single-particle spectrum
$\varepsilon (p_i<p<p_f)=\mu$. The Fermi momentum $p_F$
satisfies the condition $p_i<p_F<p_f$.} \label{fig1}
\end{figure}

To explain emergent superconductivity at $x\to x_c$, we retain
the consequences of flattening of single-particle excitation
spectra $\varepsilon({\bf p})$ (i.e. flat bands appearance) in
strongly correlated Fermi systems, see Refs.
\cite{shagrep,volovik:2015,amusia:2015} for recent reviews. At
$T=0$, the ground state of a system with a flat band is
degenerate, and the occupation numbers $n_0({\bf p})$ of
single-particle states belonging to the flat band are continuous
functions of momentum ${\bf p}$, in contrast to standard LFL
"step" from 0 to 1 at $p=p_F$, as it is seen from Fig.
\ref{fig1}. Thus at $T=0$ the superconducting order parameter
$\kappa(p)=\sqrt{n(p)(1-n(p))}\neq 0$ in the region occupied by
FC \cite{khodel:1994,shagrep,amusia:2015,qp1,qp2}. This property
is in a stark contrast to standard LFL picture, where at $T=0$
and $p=p_F$ the order parameter $\kappa(p)$ is necessarily zero,
see Fig. \ref{fig1}. Due to the fundamental difference between
the FC single-particle spectrum and that of the remainder of the
Fermi liquid, a system having FC is, in fact, a two-component
system, separated from ordinary Fermi liquid by the topological
phase transition \cite{volovik,khodel:1994,volovik:2015}. The
range $L$ of momentum space adjacent to $\mu$ where FC resides
is given by $L\simeq p_f-p_i$, see Fig. \ref{fig1}.

\section{Green functions and superfluid density}

To analyze the above emergent superconductivity quantitatively,
it is convenient to use the formalism of Gor'kov equations for
Green's functions of a superconductor
\cite{shagrep,landau9,gorkov}. For the 2D case of interest, the
solutions of Gor'kov equations \cite{shagrep,landau9,gorkov}
determine the Green's functions $F^+({\bf p},\omega)$ and
$G({\bf p},\omega)$ of a superconductor:
\begin{eqnarray}
\nonumber F^+({\bf p},\omega)
&=&\frac{-g_0\Xi^*}{(\omega -E({\bf p})+i\,0)(\omega +E({\bf p})-i\,0)};\\
G({\bf p},\omega)&=&\frac{u^2({\bf p})}{\omega -E({\bf
p})+i\,0}+\frac{v^2({\bf p})}{\omega +E({\bf
p})-i\,0}.\label{zui2}
\end{eqnarray}
Here the single-particle spectrum $\varepsilon ({\bf p})$ is
determined by Eq. \eqref{var}, and
\begin{equation}E({\bf p})=\sqrt{\xi^2({\bf
p})+\Delta^2({\bf p})};\,\,\, \frac{\Delta({\bf p})}{E({\bf
p})}=2\kappa({\bf p}), \label{SC4}\end{equation} with $\xi({\bf
p})=\varepsilon({\bf p})-\mu$. The gap  $\Delta$ and the
function $\Xi$ are given by
\begin{equation}\label{zui3}
\Delta=g_0|\Xi|,\quad i\Xi= \int F^+({\bf
p},\omega)\frac{d\omega d{\bf p} }{(2\pi)^3}.
\end{equation}
Here $g_0$ is the superconducting coupling constant. We remember
that the function $F^+({\bf p},\omega)$ has the meaning of the
wave function of Cooper pairs and $\Xi$ is the wave function of
the motion of these pairs as a whole. Taking Eqs. \eqref{SC4}
and \eqref{zui3} into account, we can rewrite Eqs. \eqref{zui2}
as
\begin{eqnarray}
F^+({\bf p},\omega)&=&-\frac{\kappa({\bf p})}{\omega -E({\bf
p})+i\,0}+\frac{\kappa({\bf p})}{\omega +E({\bf p})-i\,0}, \nonumber \\
G({\bf p},\omega)&=&\frac{u^2({\bf p})}{\omega -E({\bf
p})+i\,0}+\frac{v^2({\bf p})}{\omega +E({\bf
p})-i\,0}.\label{zui8}
\end{eqnarray}
In the case $g_0\to 0$, the gap $\Delta \to 0$, but $\Xi$ and
$\kappa({\bf p})$ remain finite if the spectrum becomes flat,
$E({\bf p})=0$, and in the interval $p_i\leq p\leq p_f$ Eqs.
(\ref{zui8}) become \cite{amusia:2015,shagrep,shagstep}
\begin{eqnarray}
F^+({\bf p},\omega )&=&-\kappa({\bf p})\left[
\frac{1}{\omega +i\,0} -\frac{1}{\omega -i\,0}\right], \nonumber  \\
G({\bf p},\omega )&=&\frac{u^2({\bf p})}{\omega
+i\,0}+\frac{v^2({\bf p})}{\omega -i\,0}.\label{zui9}
\end{eqnarray}
The parameters $v({\bf p})$ and $u({\bf p})$ are the
coefficients of corresponding Bogolubov transformation
\cite{landau9,gorkov}, $u^2({\bf p})=1-n({\bf p})$, $v^2({\bf
p})=n({\bf p})$. They are determined by the condition that the
spectrum should be flat: $\varepsilon({\bf p})=\mu$. It follows
from Eqs. \eqref{SC4} and \eqref{zui3} that
\begin{equation}\label{zui7}
i\Xi=\int F^+({\bf p},\omega )\frac{d\omega d{\bf
p}}{(2\pi)^3}=i\int\kappa({\bf p})\frac{d{\bf
p}}{(2\pi)^2}\simeq n_{FC},
\end{equation}
where $n_{FC}$ is the density of superconducting electrons,
forming the FC component, see Fig. \ref{fig1}.

We construct the functions $F^+({\bf p},\omega)$ and  $G({\bf
p},\omega)$ in the case where the constant $g_0$ is finite but
small, such that $v({\bf p})$ and $\kappa({\bf p})$ can be found
from the FC solutions of Eq. (\ref{var}). Then $\Xi$, $\Delta$
and $E({\bf p})$ are given by Eqs. (\ref{zui7}), (\ref{zui3})
and (\ref{SC4}), respectively. Substituting the functions
constructed in this manner into (\ref{zui8}), we obtain
$F^+({\bf p},\omega)$ and $G({\bf p},\omega)$. We note that Eqs.
\eqref{zui3} and \eqref{zui7} imply that the gap $\Delta$ is a
linear function of both $g_0$ and $n_{FC}$. Since $T_c\sim
\Delta$, we conclude that $T_c\propto n_{FC}\propto\rho_s$. Note
that since we  consider the overdoped HTSC case and FCQPT takes
place at $x=x_c$, $n_{FC}\propto p_F(p_f-p_i)\propto x_c-x$ with
$(p_f-p_i)/p_F\ll1$ \cite{qp1,qp2,shagrep}; therefore
\begin{equation}
n_{FC}=n_s\ll n_{el}.\label{nfc}
\end{equation}
Increasing $g_0$ causes $\Delta$ to become finite, leading to a
finite value of the effective mass $m^*_{FC}$ in the FC state
\cite{shagrep}:
\begin{equation}
m^*_{FC}\simeq
p_F\frac{p_f-p_i}{2\Delta}.\label{SC7}
\end{equation}
An important fact is to be noted here. Namely, it have been
shown in Refs \cite{amusia:2015,shagrep}, that in the FC
formalism, the BCS relations remain valid if we use the spectrum
given be Eq. \eqref{SC7}. Thus, we can use the standard BCS
approximation with the momentum independence of superconducting
coupling constant $g_0$ in the region
$|\varepsilon({\bf p})-\mu|\leq \omega_D$ so that the
interaction is supposed to be zero outside this region. Here
$\omega_D$ is a characteristic energy, proportional to the Debye
temperature. Under these suppositions, the superconducting gap
depends only on temperature and is determined by the equation
\cite{amusia:2015,khodel:1994,shagrep}

\begin{figure} [! ht]
\begin{center}
\includegraphics [width=0.52\textwidth]{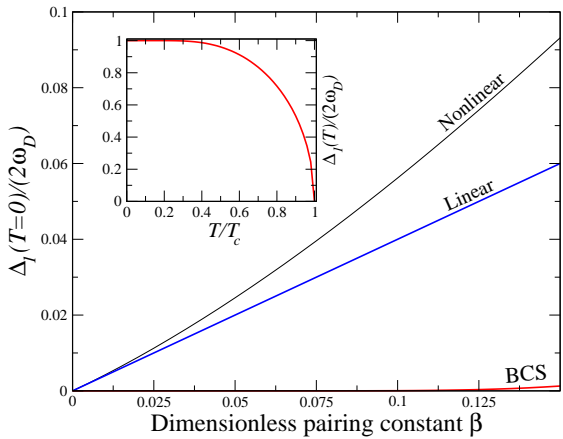}
\end{center}
\caption {(Color online) The solution of equation \eqref{gap2}
in the form $\delta(\beta)$ at $B=0.4$. Curve, marked
"Nonlinear" is a direct numerical solution of transcendental
equation \eqref{gap2}. Curve, marked "Linear" is a linear
dependence $\delta=B \beta$, while that marked "BSC" is a BCS
dependence resulting from Eq. \eqref{gap2} at $B=0$. It is seen
that FC approach permits to obtain much higher $T_c$
(proportional to superconducting gap $\Delta(T=0)$) than BCS
one. Inset reports the temperature dependence of superconducting
gap in FC approach at $B=0.4$.} \label{fig2}
\end{figure}

\begin{eqnarray}
\frac{1}{g_0}=N_{FC}\int_0^{E_0/2}\frac{d\xi}{f(\xi,\Delta)}
\tanh\frac{f(\xi,\Delta)}{2T}+\nonumber \\
+N_L\int_{E_0/2}^{\omega_D}\frac{d\xi}
{f(\xi,\Delta)}\tanh\frac{f(\xi,\Delta)}{2T}, \label{gap1}
\end{eqnarray}
where $f(\xi,\Delta)=\sqrt{\xi^2+\Delta^2(T)}$ and
$E_0=\varepsilon(p_f)-\varepsilon(p_i) \approx 2\Delta (T=0)$ is
a characteristic energy scale. Also, $N_{FC}=(p_f-p_F)p_F/(2\pi
\Delta(T=0))$ and $N_L=m^*_L/(2\pi)$ ($m^*_L$ is the effective
mass of electron of the LFL component, see Fig. \ref{fig1}) are
the densities of states of FC and non-FC electrons respectively.
In the opposite case $T=0$, as usual, $\tanh(f/(2T))=1$ and the
remaining integrals can be evaluated exactly. This yields
following equation relating the value $\Delta (T=0)$ with
superconducting coupling constant $g_0$
\begin{equation}\label{gap2}
\frac{\delta}{\beta}=B-\delta \ln \delta,
\end{equation}
where $\beta=g_0m^*_L/(2\pi)$ is dimensionless coupling
constant, $\delta=\Delta(T=0)/(2\omega_D)$ and
$B=(E_F/\omega_D)((p_f-p_F)/p_F)\ln(1+\sqrt{2})$. It is seen
that parameter $B$ depends on the width of FC interval so that
at $B=0$ ($p_f=p_F$) system is out of FC and hence is in a pure
BCS state. In this case the solution of \eqref{gap2} has
standard BCS form $\delta_{BCS}=\exp(-1/\beta)$, while at small
$\beta$ and $B \neq 0$ we obtain the linear relation between
coupling constant and gap $\delta=B\beta$, which not only
differs drastically from the BCS result, but provides much
higher $T_c$, which is directly proportional to $\Delta(T=0)$
(in the FC case $T_c \approx \Delta(T=0)/2$)
\cite{amusia:2015}. In the case of $\beta \sim 0.3$,  Fig.
\ref{fig2} portrays the solutions of equation \eqref{gap2} for
$B=0.4$ and even smaller $\beta <0.15$. It is seen that already
linear regime provides much higher $T_c$ than BSC case, while
nonlinear one comprising the complete numerical solution of Eq.
\eqref{gap2} yields even higher $T_c$. This means that FC
approach is well capable to explain the high-$T_c$
superconductivity. Inset to Fig. \ref{fig2} reports the
dependence \eqref{gap1} in dimensionless units. This dependence
is not peculiar to FC approach as it is qualitatively similar to
BCS case. In this case, the variation of "FC-parameter" $B$ (and
even putting $B=0$) does not change the situation qualitatively.

Now we analyze the superfluid density $n_s$ for finite $g_0$. As
seen from Eqs. \eqref{zui7} and \eqref{zui9}, $n_s$ emerges when
$x\simeq x_c$, and occupies the region $p_i\leq p\leq p_f$, so
that we denote $n_s=n_{FC}\propto x_c-x$, where $n_{FC}$ is the
electron density in FC phase.  As a result, we have that in
latter phase $n_s\ll n_{el}=n_{FC}+n_L$, with $n_{el}$ and $n_L$
being, respectively, the total density of electrons and that out
of FC phase. Note that the result $n_s \sim n_{el}$ does not
only follow from BCS theory of superconductivity, but is much
deeper and is pertinent to almost any superfluid system, being
the result of the Leggett theorem \cite{leg}. The short
statement of latter theorem \cite{leg} is that at $T=0$ in any
superfluid liquid $n_s \sim n_{el}$, here $n_{el}$ denotes the
number density of the liquid particles. For this theorem to be
true, however, the system should be $\rm T$ - invariant, where
$\rm T$ relates to time reversal. Since FC state, being highly
topologically nontrivial \cite{amusia:2015, baras, tun},
violates primarily the time reversal symmetry (actually it also
violates the $\rm CP$ invariance, where $\rm C$ is charge
conjugation and $\rm P$ is translation invariance, see Refs.
\cite{amusia:2015,shagrep,baras} for more details), the
inequality $n_s\ll n_{el}$ is inherent in it, as it is seen from
Eqs. \eqref{zui7} and \eqref{nfc}. This implies that the main
contribution to the above superconductivity comes from the FC
state. We conclude that in the FC case the emerging
two-component system violates the BCS condition that $n_s\simeq
n_{el}$.

\section{Penetration depth and general properties}

Now we find out if our superconductor belongs to the London
type. For that, we write down London's electrodynamics
equations: $\nabla \times {\bf j}_s=-(n_se^2/m^*){\bf B} \equiv
-(n_{FC}e^2/m^*_{FC}){\bf B}$ and $\nabla \times {\bf B} =4\pi
{\bf j}_s$, where ${\bf j}_s$ is a superconducting current.
These equations imply that the penetration depth
\begin{equation}\label{lamb}
\lambda^2=\frac{m^*_{FC}}{4\pi e^2n_{FC}}.
\end{equation}
Comparing the penetration depth \eqref{lamb} with the coherence
length $\xi_0\sim p_F/(m^*_{FC}\Delta)$, we conclude that
$\lambda >> \xi_0$ as the FC quasiparticle effective mass is
huge \cite{amusia:2015}. Thus, the superconductors are indeed of
the London type.

\begin{figure} [! ht]
\begin{center}
\includegraphics [width=0.5\textwidth]{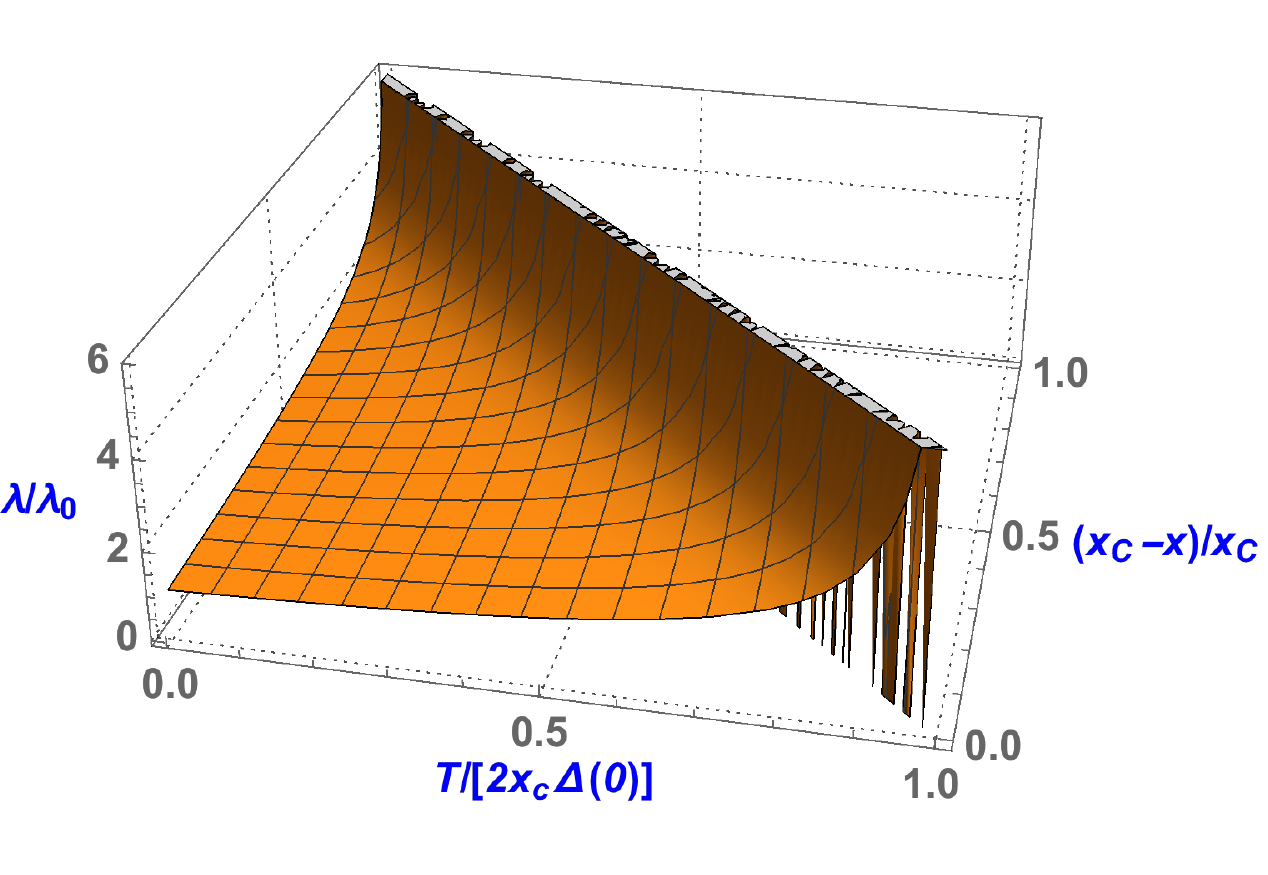}
\end{center}
\caption {(Color online) The dependence of dimensionless
penetration depth $\lambda/\lambda_0$ \eqref{lamb1} on
temperature and doping.} \label{fig3}
\end{figure}

It turns out that in FC phase, the penetration depth is a
function not only of temperature but also of doping degree $x$.
Then, it follows from Ginzburg-Landau theory, that the density
of superconducting electrons $n_s\sim T_c-T$. On the other hand,
as it has been discussed in the paper \cite{bosovic}, the
pressure enhances $n_s$, i.e. the density $x$ of charge carriers
is important. Also, it has been shown (see, e.g. Refs.
\cite{amusia:2015,shagrep}) that in superconducting phase with
FC  $T_c\simeq 2\Delta(T=0)$. This permits to use the relation
\eqref{lamb} to plot the penetration depth as a function of
temperature and doping in the form
\begin{equation}\label{lamb1}
\frac{\lambda}{\lambda_0}=\frac{1}{\sqrt{1-y-\tau}},
\end{equation}
where $y=(x_c-x)/x_c$, $\tau=T/(2\Delta(T=0)x_c)$ and
$\lambda_0$ combines all proportionality coefficients entering
the problem. The dependence \eqref{lamb1} is depicted in Fig.
\ref{fig3}. Very good qualitative agreement with experimental
data (Fig. 2a from Ref. \cite{bosovic}) is seen. Namely, doping
dependent penetration depth $\lambda$ becomes infinite at the
superconducting phase transition temperature. At zero
temperature the divergence of $\lambda$ occurs at $x\simeq x_c$,
corresponding to FC phase emergence, i.e. at $T=0$ both
superconductivity and FC phase arise. At the same time, at
higher temperatures, $\lambda$ diverges in the region $x<x_c$,
i.e. deeply inside the FC phase. This shows the "traces" of FC
at finite temperatures. This demonstrates, in turn, that our
approach, based on a concept of topological FC quantum phase
transition, describes all the essential and puzzling features of
overdoped HTSC. The main input of our model is that in two
component system with FC, occupying a small fraction of the
Fermi sphere, $n_s\simeq n_{FC}$ is much less than the total
density of the electrons. The latter also allows to verify the
validity of the well-known Uemura's law \cite{uemura} in our
case. Indeed, since $T_c\propto n_s/m^*\equiv n_{FC}/m^*_{FC}$,
we get from Eqs. \eqref{SC7} and \eqref{lamb}
\begin{equation}
\frac{\rho_s}{A}=\lambda^{-2}\simeq\frac {n_s}{m^*}\simeq \frac
{n_{FC}}{m^*_{FC}}\simeq 2\Delta\simeq T_c.\label{SC8}
\end{equation}
Taking into account that $n_{FC}\propto x_c-x$, we see that Eq.
\eqref{SC8} reproduces the main results of our paper, being in
good agreement with experimental data \cite{bosovic,zaanen}. It
is seen that the dependence of $T_c$ on $\rho_s$ is linear,
representing the observed scaling law, while $T_c$ is primary
controlled by $n_s$ \cite{bosovic}. We note that the results for
underdoped HTSC \cite{uemura,uem_n} are similar to those for
overdoped HTSC, thus being suggestive for underdoped vs
overdoped symmetry \cite{bosovic}. As a result, we observe good
agreement with the Uemura's law in overdoped LSCO as well
\cite{bosovic}.

We observe that at the doping levels $x>x_c$, where FCQPT does
not yet occur, the system is in LFL phase with resistivity
$\rho\propto T^2$, which is "more metallic" than that exhibited
in the FC phase \cite{bosovic,pagl,khod:2015,jetp2003,arch}. In
the latter phase, the superconductivity appears since FC
strongly facilitates the superconducting state. In the normal
phase, $T>T_c$, FC causes the linear $T$ dependence of
resistivity, $\rho(T)\propto T$ \cite{khod:2015,arch,qp1,qp2},
which is in good qualitative agreement with the experimental
data on LSCO and $\rm La_{2-x}Ce_xCuO_4$ \cite{bosovic,pagl}. We
note that in the transition region $x\simeq x_c$ one observes
$\rho(T)\propto T^{\alpha}$ with $\alpha\sim 1.0-2.0$
\cite{pagl,khod:2015,arch}.\\

\section{Conclusions}
In summary, we have shown that the main physical mechanism,
responsible for the unusual properties of the overdoped $\rm
La_{2-x}Sr_xCuO_4$, is the topological quantum phase transition
with the emergence of the fermion condensation. This observation
can open avenue for chemical preparation of high-$T_c$ materials
with $T_c$ up to room temperatures. We have concluded our study
of exemplifications of the new state of matter reached by
fermion condensation with an exploration of high-$T_c$
superconductors as potential hosts of fermion condensates.  In
fact, we have shown that the underlying physical mechanism
responsible for the unusual properties of the overdoped compound
$\rm La_{2-x}Sr_xCuO_4$ (LSCO) observed recently
\cite{bosovic,zaanen} may very well involve a topological
quantum phase transition that induces fermion condensation.
Since the topological FC state violates time-reversal symmetry,
the Leggett theorem no longer applies. Instead, we have
demonstrated explicitly that the superfluid number density $n_s$
turns out to be small compared to the total number density of
electrons. We have also shown that the critical temperature
$T_c$ is a linear function of $n_s$, while $n_s(T)\propto
T_c-T$. Pairing with such unusual properties is as a shadow of
fermion condensation -- a situation foretold by an exactly
solvable model \cite{qp2} long before the experimental
observations were obtained by Bo\^zovi\'c et al. \cite{bosovic}
and demonstrating that both the gap and the order parameter
exist only in the region occupied by fermion condensate. Thus,
the experimental observations \cite{bosovic} can be viewed as a
direct experimental manifestation of FC. Additionally, we have
demonstrated that at $T>T_c$ the resistivity $\rho(T)$ varies
linearly with temperature, while for $x>x_c$ it exhibits
metallic behavior, $\rho(T)\propto T^2$. Thus, pursuit of a
superconductivity formalism adapted to the presence of a fermion
condensate captures all the essential physics of overdoped LSCO
and successfully explains its most puzzling experimental
features, thereby allowing us to close the colossal gap existing
between the experiments and Bardeen-Cooper-Schrieffer-like
theories. Indeed, these findings are applicable not only to LSCO
but also for any overdoped high-temperature superconductor.

\begin{acknowledgements}
We are grateful to V.A. Khodel for valuable discussions. This
work was partly supported by U.S. DOE, Division of Chemical
Sciences, Office of Basic Energy Sciences, Office of Energy
Research.
\end{acknowledgements}


\begin{thebibliography}{99}

\bibitem{bosovic} J.I. Bo\^zovi\'c, X. He, J. Wu, and A. T. Bollinger,
Nature {\bf 536}, 309 (2016).

\bibitem{zaanen} J. Zaanen, Nature {\bf
536}, 282 (2016).

\bibitem{uemura} Y. J. Uemura {\it et al.}, \prl {\bf 62}, 2317
(1989).

\bibitem{uem_n} Y. J. Uemura {\it et al.}, Nature 364, 605 (1993).

\bibitem{bern} C. Bernhard, Ch. Niedermayer,
U. Binninger, A. Hofer, Ch. Wenger, J. L. Tallon, G. V. M.
Williams, E. J. Ansaldo, J. I. Budnick, C. E. Stronach, D. R.
Noakes, and M. A. Blankson-Mills, \prb {\bf 52}, 10488 (1995).

\bibitem{rour} P. Rourke {\it et al.}, Nat. Phys. {\bf 7}, 455 (2011).

\bibitem{bardeen} J. Bardeen, L. N. Cooper, and J. R. Schrieffer, Phys.
Rev. {\bf 106}, 162 (1957).

\bibitem{khs}  V. A. Khodel and V. R. Shaginyan, JETP Lett.
{\bf 51}, 553 (1990).

\bibitem{amusia:2015}
M. Ya. Amusia, K. G. Popov, V. R. Shaginyan, and V. A.
Stephanovich, \emph{Theory of Heavy-Fermion Compounds}, Springer
Series in Solid-State Sciences, {\bf 182}, Springer, Heidelberg,
New York, Dordrecht, London, (2015).

\bibitem{volovik:91} G. E. Volovik, JETP Lett. {\bf 53}, 222 (1991).

\bibitem{volovik} G. E. Volovik, Lect. Notes in Physics {\bf 718},
31 (2007).

\bibitem{volovik:2015} G. E. Volovik,
Phys. Scr. {\bf T164}, 014014 (2015).

\bibitem{khodel:1994}
V. A. Khodel, V. R. Shaginyan, and V. V. Khodel, Phys. Rep. {\bf
249}, 1 (1994)

\bibitem{shagrep} V. R. Shaginyan, M. Ya. Amusia, A. Z. Msezane, and K. G.
Popov, Phys. Rep. {\bf 492}, 31 (2010).

\bibitem{abrikosov} A. A. Abrikosov, J. C. Campuzano, and K.
Gofron, Physica C {\bf 214}, 73 (1993).


\bibitem{abrikosov1} A. A. Abrikosov,
Int. J. Mod. Phys. B {\bf 13}, 3405 (1999).

\bibitem{dnf} B.A. Dubrovin, A.T. Fomenko, S.P. Novikov,
{\it Modern Geometry - Methods and Applications}, (Springer, New
York, 1992).

\bibitem{vm77} G.E. Volovik, V.P. Mineev Sov. Phys. JETP {\bf 45}, 1186 (1977).

\bibitem{qp2} J. Dukelsky, V. Khodel, P. Schuck, and V. Shaginyan, Z. Phys. 102,
245 (1997)

\bibitem{pagl} K. Jin, N. P. Butch, K. Kirshenbaum, J. Paglione,
and R. L. Greene, Nature, {\bf 476}, 73 (2011).

\bibitem{khod:2015} V. A. Khodel, J. W. Clark, K. G. Popov,
and V. R. Shaginyan, JETP Lett. {\bf 101}, 413 (2015).

\bibitem{lifshitz} I. M. Lifshitz, Sov. Phys. JETP {\bf 11}, 1130 (1960).

\bibitem{ybalb} V. R. Shaginyan, A. Z. Msezane, K. G. Popov,
J. W. Clark, V. A. Khodel, and M. V. Zverev, \prb {\bf 93},
205126 (2016).

\bibitem{qp1} V.A. Khodel,  V.R. Shaginyan, and P. Schuck, JETP Lett. {\bf 63},
752 (1996)

\bibitem{landau9} E.M. Lifshits, L.P. Pitaevski
{\it Statistical Physics. Part 2}
(Butterworth-Heinemann, Oxford, 2002).

\bibitem{gorkov} A. A. Abrikosov, L. P. Gor'kov, and I. E.  Dzyaloshinski,
{\it Methods of Quantum Field Theory in Statistical Physics},
(Dover, New York, 1975).

\bibitem{shagstep} V.R. Shaginyan, A.Z. Msezane, V.A. Stephanovich,
and E.V. Kirichenko, Europhys. Lett. {\bf 76}, 898 (2006).

\bibitem{leg} A.J. Leggett, J. Stat. Phys. {\bf 93}, 927 (1998).

\bibitem{baras} V. R. Shaginyan, G. S. Japaridze, M. Ya. Amusia,
A. Z. Msezane and K. G. Popov, Europhys. Lett. {\bf 94}, 69001 (2011).

\bibitem{tun} V.R. Shaginyan, K.G. Popov, V.A. Stephanovich,
and E.V. Kirichenko, Journal of Alloys and Compounds {\bf 442}, 29 (2007).

\bibitem{jetp2003} V. R. Shaginyan, JETP Lett. {\bf 77}, 99 (2003).

\bibitem{arch} V. R. Shaginyan, A. Z. Msezane, K. G. Popov, J. W. Clark,
M. V. Zverev, and V. A. Khodel, \prb {\bf 86}, 085147 (2012).

\end{thebibliography}
\end{document}